\documentclass{article}

\usepackage{arxiv}

\usepackage[utf8]{inputenc} % allow utf-8 input
\usepackage[T1]{fontenc}    % use 8-bit T1 fonts
\usepackage{hyperref}       % hyperlinks
\usepackage{url}            % simple URL typesetting
\usepackage{booktabs}       % professional-quality tables
\usepackage{amsfonts}       % blackboard math symbols
\usepackage{nicefrac}       % compact symbols for 1/2, etc.
\usepackage{microtype}      % microtypography
\usepackage{lipsum}
\usepackage{graphicx}
\graphicspath{ {./images/} }
\usepackage{soul}
\usepackage{float}
\usepackage{graphicx}
\usepackage{caption}
\usepackage{subcaption}
\usepackage{amsmath}

\title{Machine Learning based prediction of noncentrosymmetric crystal materials}

\author{
 Yuqi Song, Joseph Lindsay, Yong Zhao,Alireza Nasiri,Steph-Yves Louis  \\
  Department of Computer Science and Engineering\\
  University of South Carolina\\
  Columbia, SC, 29201\\
   \And
Jie Ling \\
  Department of Chemistry and Biochemistry\\
  Claflin University\\
  Orangeburg, SC, 29115\\
  \And
 Ming Hu \\
  Department of Mechanical Engineering\\
  University of South Carolina\\
  Columbia, SC, 29201\\
   \And
Jianjun Hu * \\
  Department of Computer Science and Engineering\\
  University of South Carolina\\
  Columbia, SC, 29201\\
  * Correspondence author:
 \texttt{jianjunh@cse.sc.edu}
}

\begin{document}
\maketitle
\begin{abstract}
Noncentrosymmetric materials play a critical role in many important applications such as laser technology, communication systems,quantum computing, cybersecurity, and etc. However, the experimental discovery of new noncentrosymmetric materials is extremely difficult. Here we present a machine learning model that could predict whether the composition of a potential crystalline structure would be centrosymmetric or not. By evaluating a diverse set of composition features calculated using matminer featurizer package coupled with different machine learning algorithms, we find that Random Forest Classifiers give the best performance for noncentrosymmetric material prediction, reaching an accuracy of 84.8\% when evaluated with 10 fold cross-validation on the dataset with 82,506 samples extracted from Materials Project. A random forest model trained with materials with only 3 elements gives even higher accuracy of 86.9\%. We apply our ML model to screen potential noncentrosymmetric materials from 2,000,000 hypothetical materials generated by our inverse design engine and report the top 20 candidate noncentrosymmetric materials with 2 to 4 elements and top 20 borate candidates.\end{abstract}

% keywords can be removed

\section{Introduction}
\begin{figure}[H]
	\centering
	\begin{subfigure}{.45\textwidth}
		\includegraphics[width=\textwidth]{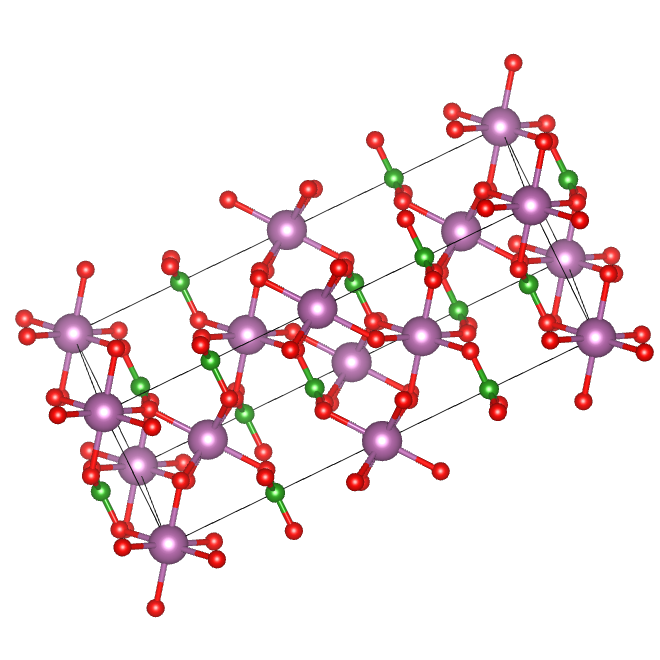}
		\caption{Centrosymmetric :ScBO3 (R$\overline{3}$c)}
	\end{subfigure}
%%%%%%%%%%%%%%
	\begin{subfigure}{.45\textwidth}
		\includegraphics[width=\textwidth]{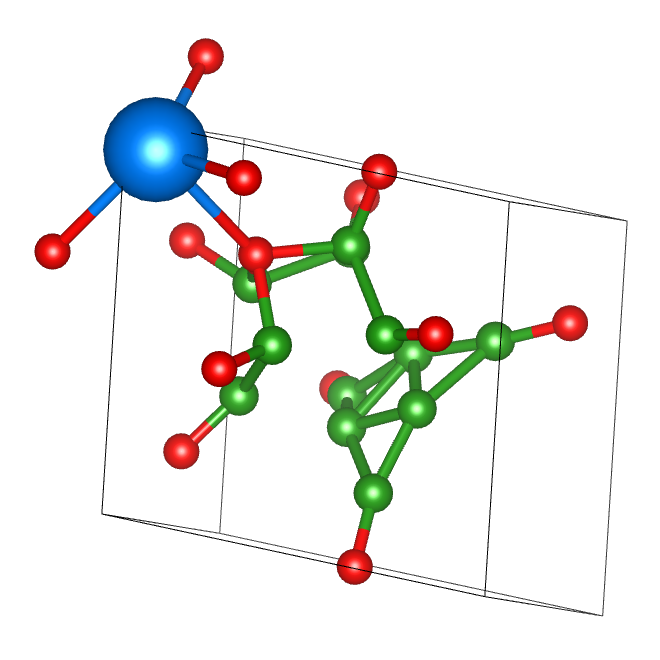}
		\caption{Noncentrosymmetric: SrB12O7 (R3)}
	\end{subfigure}
	\caption{Crystal Structures of centrosymmetric and noncentrosymmetric materials. (a) The crystal structures of ScBO3 of space group R$\overline{3}$c, where the purple nodes represent Sc atoms, the green nodes represent B atoms and red nodes are O atoms. (b) The crystal structure of SrB12O7 of space group R3, where the blue node represents Sr atom, the green nodes represent B atoms, and the red nodes are O atoms.}
    \label{example}
\end{figure}

Nonlinear optical materials (NLO), in which light waves interact with each other, are one of the key enablers for next generation of new lasers, fast telecommunication, quantum computing, quantum encryption, dynamic or optical storage data, and many other applications \cite{ok2006bulk,halasyamani1998noncentrosymmetric,kohn1999nobel,abdeldayem07}.  NLO materials are most broadly defined as those compounds capable of altering the frequency of light. Depending on the chemical and physical construct of the materials they can combine multiple photons to generate shorter wavelength photons or split one photon into several new photons of longer wavelengths. These new photons can be employed to perform all of the above applications as well as many others. The classes of NLO materials range broadly from inorganic oxides such as 
$KTiOPO_{4}$ and $LiNbO_{3}$ to semiconductors like to periodically poled GaAs, to organic polymers to metal organic framework (MOFs), and to simple small organic molecules like stilbene. This broad range of materials has many different properties and characteristics but all are united by one common factor, i.e. their lattice structure must not contain a center of symmetry and must be acentric \cite{ok2006bulk,halasyamani1998noncentrosymmetric}. This is a rigorous requirement that can only be met in well-ordered lattice structures, meaning ordered crystals. It is generally difficult to design and grow acentric single crystals and less than 15\% of all known structures are acentric. This demands exceptional determination on the part of the synthetic and crystal growth experimentalists. The process is made even more difficult by the fact that the NLO processes that enable frequency modification are inherently inefficient. Moreover, the ability to prepare new NLO materials and study their properties is not trivial and requires patient and detailed investigations. The payoff is enormous however, as the materials enable the development of devices used in next generation laser surgery, imaging, optical communication, advanced spectroscopy, optical data storage and a vast array of applications dependent on the interaction of light with matter. In Figure \ref{example}, We show the crystal structures of a centrosymmetric material and a noncentrosymmetric material, namely ScBO3 and SrB12O7.

Although the structure-property relation between NLO effects and microstructure can be used as a guide, new NLO crystals are still mainly explored using “trial and error” Edisonian approaches. A reliable determination of lattice symmetry is a crucial first step for materials characterization and analytics. Recently, a deep learning-based approach to automatically classify structures given a crystal structure (even with defects) has been recently proposed \cite{ziletti2018insightful}. Similarly, Kaufmann  et. al. \cite{kaufmann2020crystal} proposed a crystal symmetry determination method from electron diffraction using machine learning. However, these methods cannot be applied for large-scale composition based screening as they both require experimental data. On the other hand, direct numerical calculation of the optical properties of a single crystalline material from its atomic structure by accurate first-principles without any other inputs has just been made available for a few years. Studies have focused on properties such as second harmonic generation (SHG) coefficients \cite{diatta2018density} and other important optical properties such as energy band gap, refractive indices \cite{dec2018dft}, and birefringence. While first-principles calculations make it possible to predict some optical properties without any experimental data, such computation is usually tedious and very computationally demanding even for not too complicated primitive cells. Consider this: four-element compounds with different ratios can lead to a search space of 32.4 billion combinations. Currently, Density Functional Theory (DFT) based first-principles methods for optical properties calculation is out of the question for high-throughput screening of NLO materials. Especially, these methods cannot be used for discovery of new NLO materials as they all require the knowledge of the crystal structure information which is usually not available and computational prediction of crystal structures from composition is feasible only for a small subset of materials with simple compositions \cite{oganov2019structure}. In-depth understandings of the mechanism on how compositions form specific structures which further determines the NLO behavior would provide the guide for experimental explorations, and save enormous human and materials resources. On the other hand, data driven computational prediction models for noncentrosymmetric materials discovery can be used as the first step for nonlinear optical materials discovery.

In the past five years, machine learning (ML) has been increasingly applied to materials informatics problems from property prediction \cite{cao2019convolutional,hamidieh18}, to materials structure prediction, to computational screening \cite{choudhary2019accelerated}, and inverse materials design \cite{sanchez2018inverse,dan2019generative}. Among these ML algorithms and models, Random Forest (RF) models have shown great success for predicting a variety of materials properties such as the critical temperatures of superconducting materials \cite{stanev17,matsu19} and for predicting the ability of a given composition to form an amorphous ribbon of metallic glass via melt spinning \cite{ward16,ward2018machine}. In \cite{furmanchuk2016predictive}, Furmanchuk et al. utilized a RF regression model to predict the bulk modulus. RF models have also been widely used in other research areas. For example, a RF based approach showed its superiority in automatically selecting molecular descriptors for ligands of kinases and nuclear hormone receptors \cite{cano2017automatic}. On the other hand, recent years have observed tremendous success of deep learning \cite{goodfellow2016deep} based neural network models in applications such as image recognition, automatic machine translation, robotics \cite{su2020dietary}, and autonomous driving \cite{liu2017survey}. More importantly, their success in materials discovery problems such as the prediction of crystal stability \cite{ye2017crystal} and superconductor critical temperatures \cite{li2020critical}, makes it promising for other applications in materials discovery. In our previous work, we have applied machine learning and deep learning for crystal space group and crystal system prediction from composition \cite{zhao2020machine} and for formation energy prediction \cite{cao2019convolutional}.

Herein, we propose and evaluate two machine learning models including RF and multi-layer perceptron (MLP) neural network models for noncentrosymmetric classification given only material composition. The Magpie composition descriptors are used in our study. Cross-validation and hold-out experiments show that RF with Magpie features achieved the best results. A further application of our RF noncentrosymmetric prediction model to screening two million hypothetical materials generated by our generative ML model \cite{dan2019generative} allows us to identify and predict dozens of potential novel noncentrosymmetric materials with high confidence scores. 

Our contributions can be summarized as follows:

(1) We propose two machine learning algorithms (RF and MLP) for predicting noncentrosymmetric materials given only their composition.

(2) We evaluate and compare the performances of different machine learning algorithms for noncentrosymmetric materials classification.

(3) We apply our prediction models to screen the 2 million hypothetical materials generated by a generative adversarial network (GAN) based predictors and identify a list of top candidate materials with highly probable noncentrosymmetry structures. 

%%%%%%%%%%%%%%%%%%%%%%%%%%%%%%%%%%%%%%%%%%
\section{Materials and Methods}

\subsection{Feature Calculation}

To accomplish the goal of noncentrosymmetry classification, one of the key steps is to identify the most relevant features of a chemical composition that correlates with symmetry tendency of its formed structure. To do this, we have tried the myriad of featurizers provided by the matminer library \cite{ward2018matminer}, which is a Python-based software platform to facilitate data-driven methods of analyzing and predicting materials properties, such as composition, crystal structure, band structure, and more. The matminer featurizers package has a total of 5 different classes of featurizers present in the library’s current deployment, ranging from composition descriptors to structural ones. 

We use the composition featurizer's Element Property module to calculate the Magpie elemental descriptors for training our ML models. The Magpie feature set has 132 elemental descriptors \cite{ward2018machine}, composed of 6 statistics of a set of elemental properties such as atomic number in the material, space group of the material, the Magnetic Moment calculated by Density Functional Theory (DFT). Magpie feature set was selected based on our evaluations of a couple of descriptors.

\subsection{Machine learning models}

We evaluate two machine learning models for noncentrosymmetry prediction, namely, a Random Forest (RF) classifier, and a Deep Neural Network (DNN) classifier. 

Random Forest \cite{breiman2001random,liaw2002classification} is a supervised learning method that can be applied to solve classification or regression problems.
It is an ensemble algorithm that constructs a multitude of many decision trees at training time and outputs the class that is the mode of the classification of the individual trees. 
RF classifiers have shown strong prediction performance when combined with composition features in our previous studies \cite{cao19}. In our RF classifier model, we set the number of trees to be 200. This algorithm was implemented using the Scikit-Learn library in Python 3.6.

Deep learning excels at identifying patterns in unstructured data by building multiple layers to progressively extract higher-level features from the raw input to do the predictive task \cite{sze2017efficient}. For instance, Xie et al. \cite{xie2018crystal} proposed a graph convolutional neural network model for property predictions of materials and provided a universal and interpretable representation of crystalline materials. In this paper, we aimed to explore whether DNNs can achieve better predictive performance than RF models in noncentrosymmetry prediction. Therefore, we designed a MLP neural network classifier made of five fully connected layers, with four layers using LeakyReLU as their activation function and Sigmoid in the final layer for classifying. A dropout layer with a 0.05 drop rate was added to avoid overfitting. An Adam optimizer and binary cross entropy function are selected for training the DNN. In addition, the epoch, batch and learning rate are set to 50, 500, 0.001, respectively.

\subsection{Hyper-parameter tuning}

Due to various hyperparameters and the impact of their combinations on the training process and the final performance of machine learning models, manual parameter tuning is time-consuming. Hence, automatic hyperparameter tuning method is needed for finding suitable parameters. To ensure fair comparison of the ML models, we use the Bayesian optimization \cite{snoek2012practical} algorithm to find optimal hyper-parameters for RF models, which has been proven to be an effective tool. This method requires that the objective be a scalar value depending on the
hyperparamter configuration $x$, where the
maximum is sought for an expensive function $ f: \mathcal{X} \rightarrow \mathbb{R}.$

\begin{equation}
\mathbf{x}_{o p t}=\underset{\mathbf{x} \in \mathcal{X}}{\arg \max } f(\mathbf{x})
\end{equation}

We use the hyperopt package library \cite{bergstra2013hyperopt} to optimize n\_estimators, max\_depth and max\_features in RF models by supplying an optimization function which maximizes its precision.

%%%%%%%%%%%%%%%%%%%%%%%%%%%%%%%%%%%%%%%%%%
\section{Results and Discussion}

Herein, we describe the datasets, the evaluation criteria, and the experimental results. We analyze and compare the prediction performance of RF and DNN models. Besides, we discuss the application of our model to screening new hypothetical noncentrosymmetric materials. Our experiments on classifying noncentrosymmetry from composition include three parts:  cross-validation experiments, holdout experiments on Borates, and screening a two million hypothetical materials.

\subsection{Datasets}
 Crystal structures with different space groups have different centrosymmetric tendencies. It is known that there are 138 noncentrosymmetric space groups and 92 centrosymmetric space groups, the detailed space group IDs and names and their centrosymmetric property are summarized in Table \ref{table:Space group}. 

\begin{table}[H]
\caption{Space groups with noncentrosymmetric and centrosymmetric structures}
\centering
\begin{tabular}{p{75pt}p{40pt}p{300pt}}
\toprule
\textbf{} & \textbf{group IDs} & \textbf{group names} \\
\midrule
centrosymmetric	 &

2,
10-15,
47-74,
83-88,
123-142,
147-148,
162-167,
175-176,
191-194,
200-206,
221-230 &

P$\overline{1}$,
P2/m, P2\textsubscript{1}/m, C2/m, P2/c, P2\textsubscript{1}/c, C2/c,
Pmmm, Pnnn, Pccm, Pban, Pmma, Pnna, Pmna, Pcca, Pbam, Pccn, Pbcm, Pnnm, Pmmn, Pbcn, Pbca, Pnma, Cmcm, Cmca, Cmmm, Cccm, Cmma, Ccca, Fmmm, Fddd, Immm, Ibam, Ibca, Imma, 
P4/m, P4\textsubscript{2}/m, P4/n, P4\textsubscript{2}/n, I4/m, I4\textsubscript{1}/a,
P4/mmm, P4/mcc, P4/nbm, P4/nnc, P4/mbm, P4/mnc, P4/nmm, P4/ncc, P4\textsubscript{2}/mmc, P4\textsubscript{2}/mcm, P4\textsubscript{2}/nbc, P4\textsubscript{2}/nnm, P4\textsubscript{2}/mbc, P4\textsubscript{2}/mnm, P4\textsubscript{2}/nmc, P4\textsubscript{2}/ncm, I4/mmm, I4/mcm, I4\textsubscript{1}/amd, I4\textsubscript{1}/acd,
P$\overline{3}$, R$\overline{3}$,
P$\overline{3}$1m, P$\overline{3}$1c, P$\overline{3}$m1, P$\overline{3}$c1, R$\overline{3}$m, R$\overline{3}$c,
P6/m, P6\textsubscript{3}/m,
P6/mmm, P6/mcc, P6\textsubscript{3}/mcm, P6\textsubscript{3}/mmc,
Pm$\overline{3}$, Pn$\overline{3}$, Fm$\overline{3}$, Fd$\overline{3}$, Im$\overline{3}$, Pa$\overline{3}$, Ia$\overline{3}$,
Pm$\overline{3}$m, Pn$\overline{3}$n, Pm$\overline{3}$n, 
Pn$\overline{3}$m, Fm$\overline{3}$m, Fm$\overline{3}$c, Fd$\overline{3}$m, Fd$\overline{3}$c, Im$\overline{3}$m, Ia$\overline{3}$d

% & 119	   & 63,376
\\

noncentrosymmetric &

1,
3-9,
16-46,
75-82,
89-122,
143-146,
149-161,
168-174,
177-190,
195-199,
207-220 &  

P1,
P2, P2\textsubscript{1}, C2, Pm, Pc, Cm, Cc,
P222, P222\textsubscript{1}, P2\textsubscript{1}2\textsubscript{1}2, P2\textsubscript{1}2\textsubscript{1}2\textsubscript{1}, C222\textsubscript{1}, C222, F222, I222, I2\textsubscript{1}2\textsubscript{1}2\textsubscript{1}, Pmm2, Pmc2\textsubscript{1}, Pcc2, Pma2, Pca2\textsubscript{1}, Pnc2, Pmn2\textsubscript{1}, Pba2, Pna2\textsubscript{1}, Pnn2, Cmm2, Cmc2\textsubscript{1}, Ccc2, Amm2, Aem2, Ama2, Aea2, Fmm2, Fdd2, Imm2, Iba2, Ima2,
P4, P4\textsubscript{1}, P4\textsubscript{2}, P4\textsubscript{3}, I4, I4\textsubscript{1}, P$\overline{4}$, I$\overline{4}$,
 P422, P42\textsubscript{1}2, P4\textsubscript{1}22, P4\textsubscript{1}2\textsubscript{1}2, P4\textsubscript{2}22, P4\textsubscript{2}2\textsubscript{1}2, P4\textsubscript{3}22, P4\textsubscript{3}2\textsubscript{1}2, I422, I4\textsubscript{1}22,
 P4mm, P4bm, P4\textsubscript{2}cm, P4\textsubscript{2}nm, P4cc, P4nc, P4\textsubscript{2}mc, P4\textsubscript{2}bc, I4mm, I4cm, I4\textsubscript{1}md, I4\textsubscript{1}cd,
 P$\overline{4}$2m, P$\overline{4}$2c, P$\overline{4}$2\textsubscript{1}m, P$\overline{4}$2\textsubscript{1}c, P$\overline{4}$m2, P$\overline{4}$c2, P$\overline{4}$b2, P$\overline{4}$n2, I$\overline{4}$m2, I$\overline{4}$c2, I$\overline{4}$2m, I$\overline{4}$2d,
P3, P3\textsubscript{1}, P3\textsubscript{2}, R3,
P312, P321, P3\textsubscript{1}12, P3\textsubscript{1}21, P3\textsubscript{2}12, P3\textsubscript{2}21, R32, P3m1, P31m, P3c1, P31c, R3m, R3c,
P6, P6\textsubscript{1}, P6\textsubscript{5}, P6\textsubscript{2}, P6\textsubscript{4}, P6\textsubscript{3}, P$\overline{6}$,
P622, P6\textsubscript{1}22, P6\textsubscript{5}22, P6\textsubscript{2}22, P6\textsubscript{4}22, P6\textsubscript{3}22, P6mm, P6cc, P6\textsubscript{3}cm, P6\textsubscript{3}mc, P$\overline{6}$m2, P$\overline{6}$c2, P$\overline{6}$2m, P$\overline{6}$2c,
P23, F23, I23, P2\textsubscript{1}3, I2\textsubscript{1}3,
P432, P4\textsubscript{2}32, F432, F4\textsubscript{1}32, I432, P4\textsubscript{3}32, P4\textsubscript{1}32, I4\textsubscript{1}32, P$\overline{4}$3m, F$\overline{4}$3m, I$\overline{4}$3m, P$\overline{4}$3n, F$\overline{4}$3c, 
I$\overline{4}$3d

\\
\bottomrule
\label{table:Space group}
\end{tabular}
\end{table}

\begin{figure}[H]
	\centering
	\begin{subfigure}{.45\textwidth}
		\includegraphics[width=\textwidth]{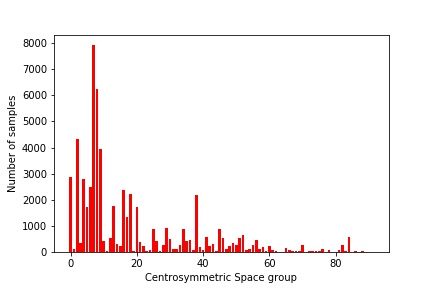}
		\caption{Centrosymmetric space groups}
	\end{subfigure}
%%%%%%%%%%%%%%
	\begin{subfigure}{.45\textwidth}
		\includegraphics[width=\textwidth]{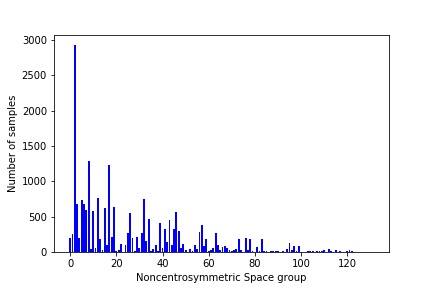}
		\caption{Non-centrosymmetric space groups}
	\end{subfigure}
	\caption{Sample distribution of noncentrosymmetric and centrosymmetric space groups in MPF dataset}
	\label{distribution of space group}
\end{figure}

We first downloaded the composition formulas of 97,217 crystal materials from the Materials Project database. We then remove those compositions belonging to multiple space groups with conflicting centrosymmetric tendencies. In total, we collecte 82,506 material compositions and assign the noncentrosymmetric property labels according to their corresponding space group. The dataset is called \textbf{MPF}, which have 60,587 positive (noncentrosymmetric) samples and 21,919 negative (centrosymmetric) samples, as shown in Table \ref{table:dataset}. The distribution of noncentrosymmetric and centrosymmetric space groups in MPF dataset are shown in Figure \ref{distribution of space group}. We find that the distribution of samples over different space groups are not well balanced.

In order to evaluate the extrapolation prediction performance of our machine learning prediction model of noncentrosymmetry, we select all the 315 borate compounds from MPF dataset and assign them as the hold-out test dataset Borates315. Borates contain boron (B) element and oxygen (O) element, which are a ubiquitous family of flame retardants found as boric acid and as a variety of salts. Previous research found that compared to other material family, borates tend to have higher percentage of nonlinear proprieties, which makes it a good hold-out test set. \cite{bubnova2017borates}. 
% To conduct hold-out evaluation of our ML model, we divide the MPF dateset into to 82190 non-borates samples and 315 borates samples, which are used as the hold-out test samples.
We further find that most borate materials include 3 elements. It is interesting to see if ML models trained with 3-element training samples can achieve better prediction performance. We select all 3-element materials from the MPF dataset and assigned them to the \textbf{MP3} dataset, which includes 30,762 centrosymmetric materials and 8,964 noncentrosymmetric materials as shown in Table \ref{table:dataset}. The motivation is to check if our classification models trained with MP3 dataset can achieve better performance when testing on the hold-out borates dataset. 

 %These materials are known as a high tendency to form noncentrosymmetric materials. There are 315 borates compounds in our hold-out test set \textbf{Borate315}. Then we remove these borates from the dataset retrieved from Materials Project database and use the remaining as the training set. 
 
%\hl{a bit confusing here:  didi MPF contains borates? did MP3 contains borate compounds?}

\begin{table}[H]
\caption{Dataset}
\centering
\begin{tabular}{cccc}
\toprule
\textbf{}	& \textbf{\#symmetry}	& \textbf{\#non symmetry}	& \textbf{\#total} \\
\midrule
 MPF   &  63,376  & 19,130  & 82,506 \\
MP3	& 30,762   & 8,964    & 39,726\\
Borates315 	& 250      & 65      & 315\\
\bottomrule
\label{table:dataset}
\end{tabular}
\end{table}

\subsection{Evaluation criteria}

To evaluate the prediction performance of our model, precision,  recall, accuracy, F1 score, and receiver operating characteristic area under the curve (ROC AUC) are used as performance metrics in this study.

The formula for these performance metrics are given as follows, where TP is number of true positives, FP is number of false positives, TN is number of true negatives, FN is number of false negatives, TPR is the true positive rate (also referred to as recall) of TP, and FPR refers to false positive rate of FP.

\begin{equation}
	Precision = \frac{TP}{TP + FP}
	\label{precison}
\end{equation}

\begin{equation}
	Recall = \frac{TP}{TP + FN}
	\label{recall}
\end{equation}

\begin{equation}
	Accuracy = \frac{TP + TN}{TP + TN + FP + FN}
	\label{accuracy}
\end{equation}

\begin{equation}
	F1-score = \frac{2TP}{2TP + FP + FN}
	\label{f1}
\end{equation}

\begin{equation}
	AUC = \int_{x=0}^{1}TPR(FPR^{-1}(x))dx
	\label{rocauc}
\end{equation}

\subsection{ Prediction performance }
To evaluate how our machine learning models can predict whether a crystal material's structure is noncentrosymmetry or not, we used two evaluation approaches: one is cross-validation over the MPF dataset and the other is the hold-out evaluation trained with non-borates datasets MPF and MP3 and tested on the Borates315 dataset. This hold-out test is especially important as the cross-validation performance can usually be over-estimated due to the redundancy of the training samples in most of the large-scale datasets such as the Materials Projects and the OQMD \cite{xiong2020evaluating}.

\subsubsection{10-fold cross-validation performance}
We set the maximum tree depth to be 20 and the number of decision trees as 200. This was later expanded to include the minimum number of samples per leaf node, the minimum number of samples required to split a node, and the maximum number of leaf nodes. With these 5 settings tuned per featurizer iteration, we then train the final prediciton RF models and make prediction, and caculate the performance scores. To further verify the performance of our RF-based models, we compare it with those of the DNN-based models. Table \ref{table:performance} shows the performances we achieved on two datasets using four evaluation criteria.

\begin{table}[H]
\caption{Ten-fold cross-validation performance of ML models for noncentrosymmetry prediction}
\centering
\begin{tabular}{llllll}
\toprule
Model     & Dataset & Precision                       & Recall                          & Accuracy                        & F1 score                        \\ \midrule
RF-based  & MPF     & 0.834                           & 0.754                           & 0.848                           & 0.781                           \\
RF-based  & MP3     &\textbf{0.845} & 0.755                           & \textbf{0.869} & \textbf{0.786} \\
DNN-based & MPF     & 0.773                           & 0.769                           & 0.785                           & 0.771                           \\
DNN-based & MP3     & 0.784                           & \textbf{0.780} & 0.792                           & 0.782                           \\ 
\toprule
\end{tabular}
\label{table:performance}
\end{table}

Firstly, we found that the precision and accuracy of the RF model are  significantly better in comparison with DNN models: the 10 fold cross-validation accuracy of RF model on the MPF dataset is 0.848 compared to 0.785, which indicates 7.89\% improvement. The F1 score of RF model is 0.781 compared to 0.771 of DNN. Although DNN achieves better Recall score, the F1-score of RF is higher than DNN's. This validates the effectiveness of our RF-based model for predicting the noncentrosymmetric property for a given material. This is consistent with a recent evaluation of different ML methods for materials property prediction \cite{robinson2019validating}. 

Secondly, comparing the results of the same RF and DNN model on the MPF dataset and the MP3 dataset, we found that each model achieved better prediction performance for the MP3 dataset. Particularly, the precision, accuracy and F1 score of the RF classifier increase to 0.845, 0.869 and 0.786, respectively.

\subsubsection{ Hold out experiment results}
\begin{figure}[H]
	\centering
	\begin{subfigure}{.45\textwidth}
		\includegraphics[width=\textwidth]{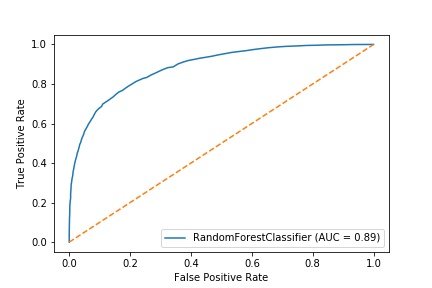}
		\caption{Cross Validation performance over MPF dataset}
		\vspace{3pt}%%yong
	\end{subfigure}
	\begin{subfigure}{.45\textwidth}
		\includegraphics[width=\textwidth]{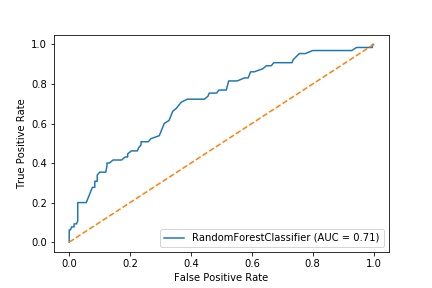}
		\caption{Holdout performance over  MPF dataset}
		\vspace{3pt}%%yong
	\end{subfigure}
	\begin{subfigure}{.45\textwidth}
		\includegraphics[width=\textwidth]{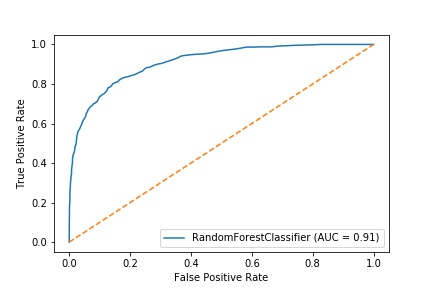}
		\caption{Cross Validation performance over MP3 dataset}
	\end{subfigure}
	\begin{subfigure}{.45\textwidth}
		\includegraphics[width=\textwidth]{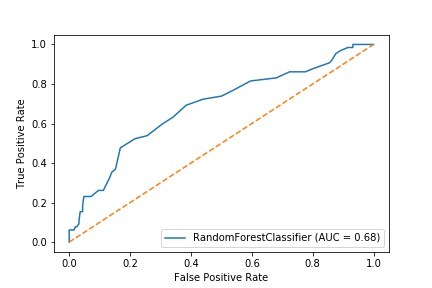}
		\caption{Holdout performance over MP3 dataset}
	\end{subfigure}
	\caption{ROC curves for cross-validation and hold-out experiments for the RF prediction models trained with the whole dataset and the 3-element dataset.}
	\label{performance}
\end{figure}

To explore the effectiveness of our model for extrapolative prediction of noncentrosymmetry where the test samples may not have the same distribution with the training set, we conducted a hold-out test over the Borates315 dataset.The training dataset is generated by filtering out all the samples of the Borates315 dataset from the MPF dataset and keeping the remaining ones, which includes 82,191 samples. Similarly, we also conduct a hold-out test for the MP3 dataset for which the training set is generated by removing all borates in the MP3 dataset. The number of samples of the no-borates 3-element training set is 39411. Their ROC curves and AUC scores are shown in Figure \ref{performance}.

In Figure \ref{performance}, each dotted yellow line corresponds to the ROC curve of a random predictor with AUC value of 0.5.  Each blue curve represents the ROC curve of the classifier. As is well known the higher value of AUC, the better performance of the classifier. Among the four sub-figures, figure (c) shows the best result, with AUC reaching 0.91. Furthermore, comparing (a) (c) with (b) (d), we can find AUC scores of cross-validation experiments are higher than those of hold-out experiments over the same two datasets, which suggests the over-estimation of model performance due to dataset sample redundancy. Meanwhile, although the performance of hold out experiments is not as good as cross validation experiments, it only uses the non-borate materials as the training data for predicting the 315 borate materials, which interprets the 0.71 and 0.68 AUC are acceptable since this is extrapolation prediction performance. Based on this analysis, we use the RF model to  predict and screen hypothetical materials from a large generated materials as discussed in detail in Section 3.4.

% \hl{needs more polishing....}

\subsubsection{The stability of our model}

To evaluate the stability of our RF model performance, we made the following Box plot, which shows that the fluctuations of precision and F1 scores for the 10-fold cross-validation experiments are less than 0.01. However, we found that the precision scores of the hold out experiments over the MPF dataset range from 0.61 to 0.67, and the F1 scores are between 0.58 to 0.64. This shows that the prediction performance of our RF models with 10-fold cross-validation experiments are more stable than those of the hold out tests.

\begin{figure}[H]
    \centering
    \includegraphics[width=15cm]{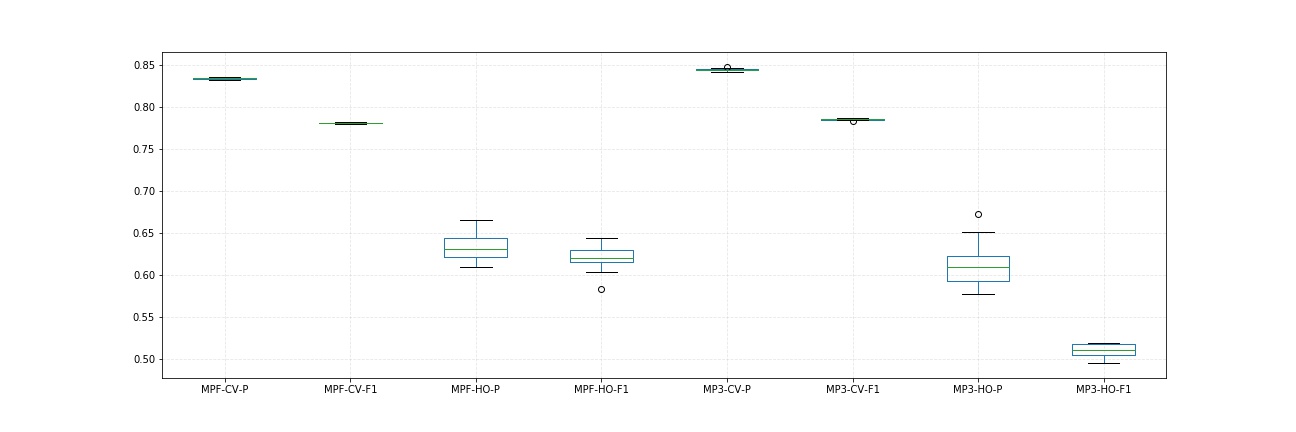}
    \caption{The stability of RF models. (MPF and MP3 are datasets; CV and H are abbreviations of cross validation and hold out; P and F1 represent Precision and F1 score respectively.)}
    \label{stability}
\end{figure}

\subsubsection{Feature importance ranking in noncentrosymmetry prediction}

\begin{figure}[H]
    \centering
    \includegraphics[width=8cm]{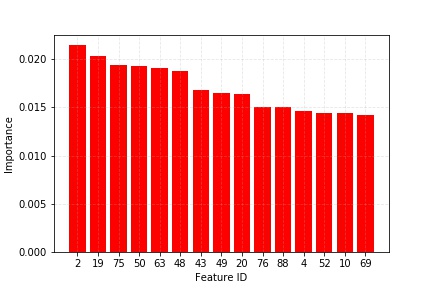}
    \caption{Ranking of top 15 features in terms of their importance scores}
    \label{importance}
\end{figure}

There are 132 descriptors in the Magpie feature set. To gain further understanding of how different descriptors affect the ML model performance, we calculated the importance scores for all descriptors in the prediction of the RF model and sorted them by their scores. The top 15 descriptors are shown in Figure \ref{importance} and the corresponding description of them are presented in Table \ref{table:importance}. 

As can be seen from Figure~\ref{importance}, the importance scores of top 15 features are above 0.014. The top six features have significantly higher scores than the remaining nine features, which shows they make more contributions to predicting the non-centrosymmetry. Combined with Table \ref{table:importance}, we find that range of atomic number, maximum melting temperature, mean number of valence, range of number of valence, mean number of Ns valence and minimum number of Nd valence are the six most important  factors. We also find that the importance of valence number to noncentrosymmetry prediction is consistent with the physical knowledge:  first the distribution of valence electrons have strong effect on chemical bond formation (strong covalent bonds or weaker ionic bonds), and thus the final crystal structure formation. Second, previous study \cite{maki1995surface} shows that the valence electrons of the atoms is involved in its nonlinear optical behavior: they construct the free electron gas, which can be polarized by the oscillating electric field and determine the harmonic excitation frequency by counting linear and nonlinear reflected waves. 
% https://sciencing.com/valence-related-bonding-behavior-atoms-8505825.html

\begin{table}[H]
\caption{Top 15 features in noncentrosymmetry prediction}
\begin{tabular}{lll}
\toprule
Feature ID & Feature Name              & Feature Description                               \\ \midrule
2          & Range Number              & Range of Atomic Number                            \\
19         & Maximum MeltingT          & Maximum Melting Temperature                       \\
75         & Mean NValence             & Mean \# Valence                         \\
50         & Range NsValence           & Range of \# Valence s-orbitals                   \\
63         & Mean NdValence            & Mean \# Valence d-orbitals                       \\
48         & Minimum NsValence         & Minimum \# Valence s-orbitals                    \\
43         & Maximum Electronegativity & Maximum Electro-negativity                        \\
49         & Maximum NsValence         & Maximum \# Valence s-orbitals                    \\
20         & Range MeltingT            & Range of Melting Temperature                      \\
76         & Avg\_dev NValence         & Mean absolute deviation of \# Valence   \\
88         & Avg\_dev NpUnfilled       & Mean absolute deviation of \# Unfilled s Orbitals \\
4          & Avg\_dev Number           & Mean absolute deviation of Atomic Number          \\
52         & Avg\_dev NsValence        & Mean absolute deviation of \# Valence s-orbitals  \\
10         & Avg\_dev MendeleevNumber  & Mean absolute deviation of Mendeleev Number       \\
69         & Mean NfValence            & Mean \# Valence f-orbitals                     \\ 
\toprule
\label{table:importance}
\end{tabular}
\end{table}

\subsection{Predicting new noncentrosymmetric materials}
To identify interesting hypothetical new NLO noncentrosymmetric materials, we applied our RF-based noncentrosymmetric materials prediction model to screen the two million hypothetical materials generated by our Generative Adversarial Network (GAN) based new materials composition generator \cite{dan2019generative}. After predicting the probability of each candidate belonging to noncentrosymmetric materials, we sort them by the probability scores and report top 20 hypothetical noncentrosymmetric materials with 2, 3 and 4 elements here in Table \ref{table:Score}. Furthermore, as we mentioned above that most borate materials are NLO materials. So we also reported top 20 borate materials with highest proability here. Please note that materials containing lanthanide and actinide elements have been filtered in these results because they are very rare.

\begin{table}[H]
\centering
\caption{Predicted hypothetical noncentrosymmetric materials with 2, 3, and 4 elements and predicted noncentrosymmetric borates (only top 20 are listed here)}
\begin{tabular}{llllllll}
\toprule
2 element &  score & 3 element & score & 4 element   & score & Borate       & Score \\  \midrule
Li4Ge     & 0.935 & AlCuSe3   & 0.960 & LaCeNdS4   & 0.975 & CB2O6        & 0.840 \\
Cu2S3     & 0.875 & Cu2AsS3   & 0.955 & LaCeNdSe4  & 0.965 & N2B4O7       & 0.715 \\
NO5       & 0.835 & Cu3As2S4  & 0.945 & CeNdEuS4   & 0.960 & CB4O6        & 0.700 \\
Li4Pb     & 0.830 & Y2CeO5    & 0.945 & CuZnInS3   & 0.955 & S3B2O8       & 0.670 \\
Li4Sn     & 0.800 & CeTb2S4   & 0.935 & AlCuZnTe4  & 0.925 & CB2O4        & 0.665 \\
Cl3S      & 0.745 & DyErC3    & 0.930 & MnNiAgSn   & 0.925 & NCB4O6       & 0.665 \\
SbC       & 0.740 & MnDy2S4   & 0.925 & AlCuInSe2  & 0.915 & CoIB4O6      & 0.660 \\
Pd2S      & 0.735 & LaSm2S4   & 0.920 & MnCoRuSn   & 0.915 & EuB4O6       & 0.655 \\
AsC       & 0.720 & ZnGaSe2   & 0.920 & LaNdUTe4   & 0.915 & ZnSnO6B4     & 0.650 \\
SeO6      & 0.715 & AlCu2Te3  & 0.915 & Cu2ZnInS6  & 0.900 & As2B2O7      & 0.635 \\
Ni3Ge2    & 0.715 & AlCu2S4   & 0.910 & NiCuSnSe3  & 0.895 & PB2O6        & 0.630 \\
Cl5S      & 0.710 & CoCd2S3   & 0.905 & MnCoAgSn   & 0.895 & ZnB2O4       & 0.625 \\
Zr2S3     & 0.695 & NbSnIr    & 0.900 & TiCoRhSn   & 0.880 & SB2O6        & 0.620 \\
S2O5      & 0.690 & NbWTe4    & 0.900 & MnFeSbO6   & 0.875 & MnZnLaEuO6B2 & 0.610 \\
LiOs      & 0.690 & VSnAu     & 0.900 & MnCu2AgS4  & 0.875 & Zn3SB2O6     & 0.600 \\
NH2       & 0.690 & CrCu2S3   & 0.895 & FeLaPbO6   & 0.875 & Sr2TaB2O6    & 0.600 \\
CrI       & 0.685 & SnTaOs    & 0.890 & V2Ni2RuSn2 & 0.875 & PbB4O6       & 0.595 \\
F3N       & 0.680 & NdDySi3   & 0.885 & MnFeBi2O6  & 0.875 & AlB2O4       & 0.585 \\
Cl6S      & 0.675 & Dy2GeS4   & 0.885 & TiCoBi2O6  & 0.870 & NbRuCl2B4O6  & 0.585 \\
S2O3      & 0.660 & Mg6MnSn   & 0.885 & SrLaNdS4   & 0.865 & C3B4O6       & 0.580 \\ \toprule
\label{table:Score}
\end{tabular}
\end{table}

As shown in Table \ref{table:Score},
the probability score range of top 20 2-element materials, 3-element materials, 4-element materials and borate materials are 0.935 to 0.660, 0.960 to 0.885, 0.975 to 0.865 and 0.885 to 0.670, respectively. It is clear that the predicted noncentrosymmetic probabilities of 3 element materials are higher than those of 2-element materials and 4-element materials. As those material are generated and hypothetical, we can only give the predicted noncentrosymmetry scores, which may guide experimental work to verify them in future research, which may further validate the effectiveness and the predictive capability of our models.
More prediction results can be provided by the corresponding author upon reasonable request.

%%%%%%%%%%%%%%%%%%%%%%%%%%%%%%%%%%%%%%%%%%
\section{Conclusions}

Computational prediction of noncentrosymmetry of a given composition can be used for fast screening new nonlinear optical materials. Here we developed and evaluated two machine learning models including a Random Forest Classifier and a neural network model for computational prediction of materials noncentrosymmetry given only their composition information. By using the Magpie composition features, our best prediction model based on Random forest can achieve an accuracy of 84.8\% when evaluated using 10-fold cross-validation over the Material Projects database. Further experiments showed that when the prediction model is trained only on 3-element samples, it can achieve even higher performance for the test set, which is made of mostly 3-element materials. A feature importance calculation shows the top six contribution factors for predicting noncentrosymmetry, many of which are related to the distribution of valence electrons. which is consistent with current physichochemical principles. Our developed model can be applied to discovering novel nonlinear materials as we conduct large-scale screening over two million hypothetical materials.

%%%%%%%%%%%%%%%%%%%%%%%%%%%%%%%%%%%%%%%%%%
% \section{Patents}
% This section is not mandatory, but may be added if there are patents resulting from the work reported in this manuscript.

%%%%%%%%%%%%%%%%%%%%%%%%%%%%%%%%%%%%%%%%%%
\vspace{6pt} 

%%%%%%%%%%%%%%%%%%%%%%%%%%%%%%%%%%%%%%%%%%
%% optional
%\supplementary{The following are available online at \linksupplementary{s1}, Figure S1: title, Table S1: title, Video S1: title.}


%%%%%%%%%%%%%%%%%%%%%%%%%%%%%%%%%%%%%%%%%%
\section{Author contributions}
conceptualization, J.H. and J.L.; methodology, J.H., S.Y., Y.Z., J.L.; software, Y.S., Y.Z., J.L.; validation, Y.S and J.L.;  investigation, J.H., Y.S., J.L.;  data curation, A.N. and J.L.; writing--original draft preparation, J.H., Y.S., and J.L.; writing--review and editing, J.H.,Y.S., J.L., M.H.; visualization, Y.S.; supervision, J.H. ; project administration, J.H.; funding acquisition, J.H., M.H. and J.L.

%%%%%%%%%%%%%%%%%%%%%%%%%%%%%%%%%%%%%%%%%%
% \funding{Please add: ``This research received no external funding'' or ``This research was funded by NAME OF FUNDER grant number XXX.'' and  and ``The APC was funded by XXX''. Check carefully that the details given are accurate and use the standard spelling of funding agency names at \url{https://search.crossref.org/funding}, any errors may affect your future funding.}

%%%%%%%%%%%%%%%%%%%%%%%%%%%%%%%%%%%%%%%%%%
\section{Acknowledgements}
Research reported in this work was supported in part by the NSF and SC EPSCoR Program under award number (NSF Award \#OIA-1655740 and SC EPSCoR grant GEAR-CRP 2019-GC02). The views, perspective, and content do not necessarily represent the official views of the SC EPSCoR Program nor those of the NSF. This work was also partially supported by NSF under grant  1940099 and 1905775.

%%%%%%%%%%%%%%%%%%%%%%%%%%%%%%%%%%%%%%%%%%
The authors declare no conflict of interest.

Data Availability:

The data required to reproduce these findings are downloaded from Materials Project database at https://materialsproject.org/. 

%%%%%%%%%%%%%%%%%%%%%%%%%%%%%%%%%%%%%%%%%%
% \nocite{*} % remove this when all references are made 
\bibliographystyle{unsrt}

% \bibliography{references}

\begin{thebibliography}{10}

\bibitem{ok2006bulk}
Kang~Min Ok, Eun~Ok Chi, and P~Shiv Halasyamani.
\newblock Bulk characterization methods for non-centrosymmetric materials:
  second-harmonic generation, piezoelectricity, pyroelectricity, and
  ferroelectricity.
\newblock {\em Chemical Society Reviews}, 35(8):710--717, 2006.

\bibitem{halasyamani1998noncentrosymmetric}
P~Shiv Halasyamani and Kenneth~R Poeppelmeier.
\newblock Noncentrosymmetric oxides.
\newblock {\em Chemistry of Materials}, 10(10):2753--2769, 1998.

\bibitem{kohn1999nobel}
Walter Kohn.
\newblock Nobel lecture: Electronic structure of matter—wave functions and
  density functionals.
\newblock {\em Reviews of Modern Physics}, 71(5):1253, 1999.

\bibitem{abdeldayem07}
Hossin~A. Abdeldayem and Donald~O. Frazier, editors.
\newblock {\em Nonlinear Optics and Applications}.
\newblock Research Signpost, 2007.

\bibitem{ziletti2018insightful}
Angelo Ziletti, Devinder Kumar, Matthias Scheffler, and Luca~M Ghiringhelli.
\newblock Insightful classification of crystal structures using deep learning.
\newblock {\em Nature communications}, 9(1):1--10, 2018.

\bibitem{kaufmann2020crystal}
Kevin Kaufmann, Chaoyi Zhu, Alexander~S Rosengarten, Daniel Maryanovsky,
  Tyler~J Harrington, Eduardo Marin, and Kenneth~S Vecchio.
\newblock Crystal symmetry determination in electron diffraction using machine
  learning.
\newblock {\em Science}, 367(6477):564--568, 2020.

\bibitem{diatta2018density}
A~Diatta, J~Rouquette, P~Armand, and P~Hermet.
\newblock Density functional theory prediction of the second harmonic
  generation and linear pockels effect in trigonal bazno2.
\newblock {\em The Journal of Physical Chemistry C}, 122(37):21277--21283,
  2018.

\bibitem{dec2018dft}
Bart{\l}omiej Dec and Robert Bogdanowicz.
\newblock Dft studies of refractive index of boron-doped diamond.
\newblock {\em Photonics Letters of Poland}, 10(2):39--41, 2018.

\bibitem{oganov2019structure}
Artem~R Oganov, Chris~J Pickard, Qiang Zhu, and Richard~J Needs.
\newblock Structure prediction drives materials discovery.
\newblock {\em Nature Reviews Materials}, 4(5):331--348, 2019.

\bibitem{cao2019convolutional}
Zhuo Cao, Yabo Dan, Zheng Xiong, Chengcheng Niu, Xiang Li, Songrong Qian, and
  Jianjun Hu.
\newblock Convolutional neural networks for crystal material property
  prediction using hybrid orbital-field matrix and magpie descriptors.
\newblock {\em Crystals}, 9(4):191, 2019.

\bibitem{hamidieh18}
Kam Hamidieh.
\newblock A data-driven statistical model for predicting the
  criticaltemperature of a superconductor.
\newblock Technical report, Statistics Department, University of Pennsylvania,
  Wharton, PA, 2018.

\bibitem{choudhary2019accelerated}
Kamal Choudhary, Marnik Bercx, Jie Jiang, Ruth Pachter, Dirk Lamoen, and
  Francesca Tavazza.
\newblock Accelerated discovery of efficient solar cell materials using quantum
  and machine-learning methods.
\newblock {\em Chemistry of Materials}, 31(15):5900--5908, 2019.

\bibitem{sanchez2018inverse}
Benjamin Sanchez-Lengeling and Al{\'a}n Aspuru-Guzik.
\newblock Inverse molecular design using machine learning: Generative models
  for matter engineering.
\newblock {\em Science}, 361(6400):360--365, 2018.

\bibitem{dan2019generative}
Yabo Dan, Yong Zhao, Xiang Li, Shaobo Li, Ming Hu, and Jianjun Hu.
\newblock Generative adversarial networks (gan) based efficient sampling of
  chemical space for inverse design of inorganic materials.
\newblock {\em arXiv preprint arXiv:1911.05020}, 2019.

\bibitem{stanev17}
Valentin Stanev, Corey Oses, A.~Kusne, Efrain Rodriguez, Johnpierre Paglione,
  Stefano Curtarolo, and I.~Takeuchi.
\newblock Machine learning modeling of superconducting critical temperature.
\newblock {\em npj Computational Materials}, 4, 09 2017.

\bibitem{matsu19}
Kaname Matsumoto and Tomoya Horide.
\newblock Acceleration search method of higher tc superconductors by machine
  learning algorithm.
\newblock {\em Applied Physics Express}, 12, 06 2019.

\bibitem{ward16}
Logan Ward.
\newblock A general-purpose machine learning framework for predicting
  properties of inorganic materials.
\newblock In Pat Langley, editor, {\em Nature News}. Nature Publishing Group,
  2016.

\bibitem{ward2018machine}
Logan Ward, Stephanie~C O'Keeffe, Joseph Stevick, Glenton~R Jelbert, Muratahan
  Aykol, and Chris Wolverton.
\newblock A machine learning approach for engineering bulk metallic glass
  alloys.
\newblock {\em Acta Materialia}, 159:102--111, 2018.

\bibitem{furmanchuk2016predictive}
Al'ona Furmanchuk, Ankit Agrawal, and Alok Choudhary.
\newblock Predictive analytics for crystalline materials: bulk modulus.
\newblock {\em RSC advances}, 6(97):95246--95251, 2016.

\bibitem{cano2017automatic}
Gaspar Cano, Jose Garcia-Rodriguez, Alberto Garcia-Garcia, Horacio
  Perez-Sanchez, J{\'o}n~Atli Benediktsson, Anil Thapa, and Alastair Barr.
\newblock Automatic selection of molecular descriptors using random forest:
  Application to drug discovery.
\newblock {\em Expert Systems with Applications}, 72:151--159, 2017.

\bibitem{goodfellow2016deep}
Ian Goodfellow, Yoshua Bengio, and Aaron Courville.
\newblock {\em Deep learning}.
\newblock MIT press, 2016.

\bibitem{su2020dietary}
Zhidong Su, Yang Li, and Guanci Yang.
\newblock Dietary composition perception algorithm using social robot audition
  for mandarin chinese.
\newblock {\em IEEE Access}, 8:8768--8782, 2020.

\bibitem{liu2017survey}
Weibo Liu, Zidong Wang, Xiaohui Liu, Nianyin Zeng, Yurong Liu, and Fuad~E
  Alsaadi.
\newblock A survey of deep neural network architectures and their applications.
\newblock {\em Neurocomputing}, 234:11--26, 2017.

\bibitem{ye2017crystal}
Weike Ye, Chi Chen, Zhenbin Wang, Iek-Heng Chu, and Shyue Ong.
\newblock Deep neural networks for accurate predictions of crystal stability.
\newblock {\em Nature Communications}, 9, 12 2017.

\bibitem{li2020critical}
Shaobo Li, Yabo Dan, Xiang Li, Tiantian Hu, Rongzhi Dong, Zhuo Cao, and Jianjun
  Hu.
\newblock Critical temperature prediction of superconductors based on atomic
  vectors and deep learning.
\newblock {\em Symmetry}, 12(2):262, 2020.

\bibitem{zhao2020machine}
Yong Zhao, Yuxin Cui, Zheng Xiong, Jing Jin, Zhonghao Liu, Rongzhi Dong, and
  Jianjun Hu.
\newblock Machine learning-based prediction of crystal systems and space groups
  from inorganic materials compositions.
\newblock {\em ACS Omega}, 2020.

\bibitem{ward2018matminer}
Logan Ward, Alexander Dunn, Alireza Faghaninia, Nils~ER Zimmermann, Saurabh
  Bajaj, Qi~Wang, Joseph Montoya, Jiming Chen, Kyle Bystrom, Maxwell Dylla,
  et~al.
\newblock Matminer: An open source toolkit for materials data mining.
\newblock {\em Computational Materials Science}, 152:60--69, 2018.

\bibitem{breiman2001random}
Leo Breiman.
\newblock Random forests.
\newblock {\em Machine learning}, 45(1):5--32, 2001.

\bibitem{liaw2002classification}
Andy Liaw, Matthew Wiener, et~al.
\newblock Classification and regression by randomforest.
\newblock {\em R news}, 2(3):18--22, 2002.

\bibitem{cao19}
Zhuo Cao, Yabo Dan, Zheng Xiong, Chengcheng Niu, Xiang Li, Songrong Qian, and
  Jianjun Hu.
\newblock Convolutional neural networks for crystal material property
  prediction using hybrid orbital-field matrix and magpie descriptors.
\newblock {\em Crystals}, 9:191, 04 2019.

\bibitem{sze2017efficient}
Vivienne Sze, Yu-Hsin Chen, Tien-Ju Yang, and Joel~S Emer.
\newblock Efficient processing of deep neural networks: A tutorial and survey.
\newblock {\em Proceedings of the IEEE}, 105(12):2295--2329, 2017.

\bibitem{xie2018crystal}
Tian Xie and Jeffrey~C Grossman.
\newblock Crystal graph convolutional neural networks for an accurate and
  interpretable prediction of material properties.
\newblock {\em Physical review letters}, 120(14):145301, 2018.

\bibitem{snoek2012practical}
Jasper Snoek, Hugo Larochelle, and Ryan~P Adams.
\newblock Practical bayesian optimization of machine learning algorithms.
\newblock In {\em Advances in neural information processing systems}, pages
  2951--2959, 2012.

\bibitem{bergstra2013hyperopt}
James Bergstra, Dan Yamins, and David~D Cox.
\newblock Hyperopt: A python library for optimizing the hyperparameters of
  machine learning algorithms.
\newblock In {\em Proceedings of the 12th Python in science conference}, pages
  13--20. Citeseer, 2013.

\bibitem{bubnova2017borates}
Rimma Bubnova, Sergey Volkov, Barbara Albert, and Stanislav Filatov.
\newblock Borates—crystal structures of prospective nonlinear optical
  materials: High anisotropy of the thermal expansion caused by anharmonic
  atomic vibrations.
\newblock {\em Crystals}, 7(3):93, 2017.

\bibitem{xiong2020evaluating}
Zheng Xiong, Yuxin Cui, Zhonghao Liu, Yong Zhao, Ming Hu, and Jianjun Hu.
\newblock Evaluating explorative prediction power of machine learning
  algorithms for materials discovery using k-fold forward cross-validation.
\newblock {\em Computational Materials Science}, 171:109203, 2020.

\bibitem{robinson2019validating}
Matthew~C Robinson, Robert~C Glen, and Alpha~A Lee.
\newblock Validating the validation: Reanalyzing a large-scale comparison of
  deep learning and machine learning models for bioactivity prediction.
\newblock {\em arXiv preprint arXiv:1905.11681}, 2019.

\bibitem{maki1995surface}
Jeffery~J Maki, Martti Kauranen, and Andr{\'e} Persoons.
\newblock Surface second-harmonic generation from chiral materials.
\newblock {\em Physical review B}, 51(3):1425, 1995.

\end{thebibliography}

\end{document}